\documentclass[11pt]{amsproc}

\theoremstyle{definition}

\theoremstyle{remark}

\numberwithin{equation}{section}

\usepackage[all]{xy}
\usepackage{graphicx}
\usepackage{hyperref}
\usepackage{color}
\usepackage[mathscr]{eucal}
\usepackage{slashbox}
\def\clap#1{\hbox to 0pt{\hss#1\hss}}
\def\mathrlapinternal#1#2{%
\rlap{$\mathsurround=0pt#1{#2}$}}
\def\mathclapinternal#1#2{%
\clap{$\mathsurround=0pt#1{#2}$}}	
\def\mathrlap{\mathpalette\mathrlapinternal}
\def\mathclap{\mathpalette\mathclapinternal}
\newcommand{\Q}{\ensuremath{{\mathbb{Q}}}}
\newcommand{\R}{\ensuremath{{\mathbb{R}}}}
\newcommand{\C}{\ensuremath{{\mathbb{C}}}}
\newcommand{\Z}{\ensuremath{{\mathbb{Z}}}}
\newcommand{\ZZZ}{\ensuremath{{\Z_3\times\Z_3}}}
\newcommand{\Xt}{{\ensuremath{\widetilde{X}}}}
\newcommand{\Ct}{{\ensuremath{\widetilde{C}}}}
\newcommand{\CP}{\ensuremath{\mathop{\null {\mathbb{P}}}}\nolimits}
\newcommand{\Rep}[1]{\ensuremath{\mathbf{\underline{#1}}}}
\DeclareMathOperator{\Li}{Li}
\newcommand{\ThetaEeight}{\ensuremath{\Theta_{E_8}}}
\newcommand{\Fprepotential}{\mathscr{F}}

\newcommand{\Fprepot}[1]{\ensuremath{\Fprepotential_{{#1},0}}}
\newcommand{\FprepotNP}[1]{\ensuremath{\Fprepot{#1}^\text{np}}}

\newcommand{\FprepotXNP}{\FprepotNP{X}}

\begin{document}

\title{Worldsheet Instantons and Torsion Curves}

\author[V.~Braun]{Volker Braun}
\address{Department of Physics, University of Pennsylvania,        
  209 S. 33rd Street, Philadelphia, PA 19104--6395, USA}
\email{vbraun@physics.upenn.edu}
\thanks{This research was supported in part by the Department of
  Physics and the Math/Physics Research Group at the University of
  Pennsylvania under cooperative research agreement DE-FG02-95ER40893
  with the U.~S.~Department of Energy and an NSF Focused Research
  Grant DMS0139799 for ``The Geometry of Superstrings'', in part by
  the Austrian Research Funds FWF grant number P18679-N16, in part by
  the European Union RTN contract MRTN-CT-2004-005104, in part by the
  Italian Ministry of University (MIUR) under the contract PRIN
  2005-023102 ``Superstringhe, brane e interazioni fondamentali'', and
  in part by the Marie Curie Grant MERG-2004-006374.
  \\ \indent
  V.~Braun would like to thank the organizers of the Sowers Workshop,
  and in particular Mark Sowers, for the opportunity to present this
  work.}

\author[M.~Kreuzer]{Maximilian Kreuzer}
\address{Institute for Theoretical Physics, Vienna University of
  Technology, Wiedner Hauptstr. 8-10, 1040 Vienna, Austria}
\email{Maximilian.Kreuzer@tuwien.ac.at}

\author[B.~A.~Ovrut]{Burt A. Ovrut}
\address{Department of Physics, University of Pennsylvania,        
  209 S. 33rd Street, Philadelphia, PA 19104--6395, USA}
\email{ovrut@physics.upenn.edu}

\author[E.~Scheidegger]{Emanuel Scheidegger} 
\address{
  Institut f\"ur Mathematik, 
  Universit\"at Augsburg, 86135 Augsburg, Germany
} 
\email{emanuel.scheidegger@math.uni-augsburg.de}

\date{May 16, 2007 and, in revised form, December 1, 2007.}


\begin{abstract}
  We study aspects of worldsheet instantons relevant to a heterotic
  standard model. The non-simply connected Calabi-Yau threefold used
  admits \ZZZ{} Wilson lines, and a more detailed investigation shows
  that the homology classes of curves are
  $H_2(X,\Z)=\Z^3\oplus(\Z_3\oplus\Z_3)$. We compute the genus-$0$
  prepotential, this is the first explicit calculation of the
  Gromov-Witten invariants of homology classes with torsion (finite
  subgroups)\cite{Braun:2007tp, Braun:2007xh, Braun:2007vy}. In
  particular, some curve classes contain only a single instanton. This
  ensures that the Beasley-Witten cancellation of instanton
  contributions cannot happen on this (non-toric) Calabi-Yau
  threefold.
\end{abstract}

\maketitle

\section{Heterotic Standard Models}
\label{sec:Models}

\subsection{Heterotic M-theory}
\label{sec:HeteroticMtheory}

Probably the most promising corner of string theory to construct
models with realistic particle spectra is heterotic M-theory, also
known as the Horava Witten setup~\cite{Horava:1996ma,
  Lukas:1998hk}. In it, the spacetime is taken to be Minkowski space
$\R^{3,1}$ times a Calabi-Yau threefold $X$ times an interval $I$ in
the eleventh direction.
\begin{figure}[htbp]
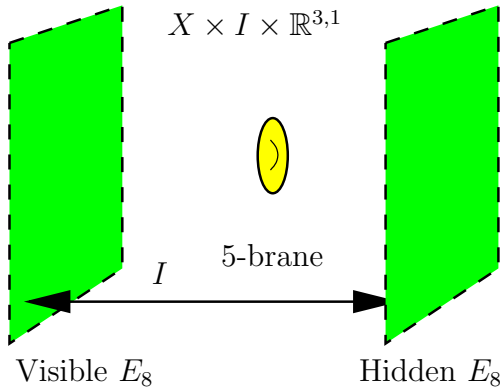
\caption{Horava-Witten setup.}
\label{fig:HoravaWitten}
\end{figure}
The two 10-dimensional boundaries each support a $E_8$ gauge theory,
one of which should be broken by instantons and/or Wilson lines to the
standard model gauge group. The other $E_8^\text{hid}$ is then hidden
and only couples gravitationally to the visible sector. In addition,
there can be a number of $5$-branes wrapping a curve of the Calabi-Yau
threefold in the interior of the interval. In order to realize
weak-scale supersymmetry breaking either the $5$-brane or the hidden
$E_8^\text{hid}$ gauge sector should break supersymmetry and
gravitationally mediate the effect to the ($N=1$ supersymmetric)
visible $E_8^\text{vis}$ sector.

\subsection{A Heterotic Standard Model}
\label{sec:HSM}

In order to find a compactification of heterotic M-theory one thus has
to specify a Calabi-Yau threefold $X$ and two $E_8$ gauge bundles on
it. For our purposes it will be convenient to pick an $SU(n)$ subgroup
in $E_8$, and use this group embedding to promote $SU(n)$ bundles
(that is, vector bundles $V_\text{vis}$ and $V_\text{hid}$) to $E_8$
bundles. The commutant of its holonomy $\mathop{\text{Hol}}( V_\text{vis} )$
inside the visible $E_8^\text{vis}$ is then the unbroken low-energy
gauge group in $4$ dimensions.

Here, we are going to be interested in the case where $V_\text{vis}$
is a rank $4$ holomorphic, slope stable (hence preserving $N=1$
by itself) vector bundle together with a $\ZZZ$ Wilson line. In other
words,
\begin{equation}
  \mathop{\text{Hol}}( V_\text{vis} ) 
  = SU(4) \times \ZZZ
  \quad \subset E_8^\text{vis} 
  .
\end{equation}
The unbroken gauge group is then
\begin{equation}
  SU(3)_C \times SU(2)_L \times U(1)_Y \times U(1)_{B-L}
  ,
\end{equation}
the standard model gauge group times a $U(1)_{B-L}$ which constrains
proton decay.

The (visible) matter fields of the effective $4$-dimensional theory
correspond to the zero modes of the Dirac operator
$~{\raisebox{0.15mm}{\slash}\hspace{-3mm} D}_{E_8}$ on the Calabi-Yau
threefold $X$. They can be computed as the $H^1$ cohomology groups of
the vector bundle $V_\text{vis}$ and associated bundles
$V_\text{vis}^\vee$, $\wedge^2 V_\text{vis}$,
$\dots$. In~\cite{Braun:2005nv, Braun:2006ae}, we constructed a
suitable vector bundle yielding precisely the MSSM matter spectrum of
\begin{itemize}
\item three families of quarks and leptons,
\item one (up, down) pair of Higgs, and
\item no other charged fields.
\end{itemize}
One interesting possibility for the hidden sector (cancelling the
heterotic anomaly) consists of an unbroken $E_8^\text{hid}$ plus
anti-five branes wrapping a rigid curve in the bulk~\cite{Braun:2006th}.
See also~\cite{Buchmuller:2005jr, Bouchard:2005ag} for other heterotic
constructions.

\subsection{Bundle Superpotential}
\label{sec:bundlesupo}

Mathematically, deformations of holomorphic vector bundles are
unobstructed (at a smooth point in the bundle moduli
space). Therefore, there cannot be any perturbative superpotential for
the bundle moduli fields. 

However, world-sheet instantons can, in principle, generate a
superpotential. The action for a worldsheet instanton wrapped on a
holomorphic curve $C$ is
\begin{equation}
  S[C\to X]
  =
  \int_C \omega
  ,
\end{equation}
where $\omega$ is the complexified K\"ahler class
\begin{equation}
  \label{eq:CKahlerClass}
  \omega = iJ + B = i t^a J_a + B
  .
\end{equation}
The instanton-generated superpotential $W_\text{bundle}$ is then the
sum over all holomorphic curves and weighted with the Pfaffian (a
function of those moduli which determine the bundle restricted to
$C$):
\begin{equation}
  \label{eq:Wbundle}
  W_\text{bundle}\big(
  \phi, 
  c, 
  t
  \big)
  = 
  \sum_{C \subset X}
  \mathop{\text{Pf}}(
  \phi, c
  )
  e^{i\int_C \omega}
  ,
\end{equation}
where
\begin{itemize}
\item $\phi$ are the $\dim H^1(X, V\otimes V^\vee)$ vector bundle moduli,
\item $c$ are the $h^{21}(X)$ complex structure moduli, and
\item $t=\{t^a\}$ are the $h^{11}(X)$ K\"ahler moduli.
\end{itemize}
Ignoring any torsion parts in homology and cohomology for the moment,
we can generate $H^2(X,\Z)$ by the basis two-forms $J_a$ of
eq.~\eqref{eq:CKahlerClass} and, therefore, write the homology class
of the curve $C$ as
\begin{equation}
  [C] = 
  (n_1,\dots, n_{h^{11}}) = \Big( \int_C J_1, \dots, \int_C J_{h^{11}}
  \Big)
  \in \Z^{h^{11}}\simeq H_2(X)
\end{equation}
Usually, we use the Fourier-transformed K\"ahler moduli $q_a =
e^{2\pi i t^a}$ and write the instanton action part as 
\begin{equation}
  e^{i\int_C t^a J_a} = \prod q_a^{n_a}
  .
\end{equation}

Unlikely as this may seem, in common constructions, like gauged linear
sigma models and monad constructions, the bundle moduli superpotential
eq.~\eqref{eq:Wbundle} vanishes. One possible explanation was offered
by Beasley and Witten, who have argued~\cite{Beasley:2003fx} that the
sum over all instantons within a given homology class vanishes under
certain circumstances. That is, rewrite eq.~\eqref{eq:Wbundle} as
\begin{equation}
  W_\text{bundle}
  =
  \sum_{C \subset X}
  \mathop{\text{Pf}}(
  \phi, c
  )
  \prod q_a^{n_a}
  =
  \sum_{\vec{n}\in H_2(X) }
  \underbrace{\left( 
      \sum_{[C] = \vec{n}}
      \mathop{\text{Pf}}(
      \phi, c
      )
    \right)}_{=0\text{ ?}}
  \prod q_a^{n_a}
\end{equation}
Then, for bundles coming from the toric ambient space, the inner
sum actually does vanish. As we will see later in
Subsection~\ref{sec:numbers}, this argument does not apply to our
vector bundle. Since our underlying Calabi-Yau threefold $X$ is not a
toric complete intersection, this should not be too surprising.

\section{The \texorpdfstring{$\mathbf{\ZZZ}$}{Z/3 x Z/3} Calabi-Yau Threefold}
\label{sec:ZZZ}

\subsection{The Manifold}
\label{sec:Manifold}

We now construct the Calabi-Yau threefold $X$ as a free
$\ZZZ$-quotient of a simply connected Calabi-Yau threefold
$\Xt$. Clearly, then, the fundamental group $\pi_1(X)=\ZZZ$ allows for
suitable Wilson lines, see Subsection~\ref{sec:HSM}. 

The covering space $\Xt$ is the complete intersection Calabi-Yau
threefold defined by a degree-$(3,1,0)$ and~$(0,1,3)$ polynomial in
$\CP^2\times\CP^1\times\CP^2$. Discrete symmetries appear for special
values of the complex structure moduli. In fact, there are \emph{two
  different} loci in the complex structure moduli space with free
$\Z_3\times\Z_3$ group actions, leading to two different
$\Z_3\times\Z_3$ quotients. The two quotients have different complex
structures, but cannot be distinguished at the level of cohomology
groups. One of the quotients was used in~\cite{Braun:2005nv,
  Braun:2006ae} to construct a heterotic MSSM, and the explicit group
action and polynomial equations can be found
in~\cite{Braun:2007sn}. However, we are now going to consider the
other quotient, where one can use toric mirror symmetry. The covering
space of the second quotient
\begin{equation}
  \Xt 
  \subset
  \CP^2_{[x_0:x_1:x_2]} \times 
  \CP^1_{[t_0:t_1]} \times 
  \CP^2_{[y_0:y_1:y_2]}
\end{equation}
is the complete intersection (see also~\cite{Candelas:2007ac})
\begin{equation}
  \begin{array}{c}
    t_0 \Big( x_0^3+x_1^3+x_2^3 \Big) + 
    t_1 \Big( x_0 x_1 x_2 \Big)
    = 0 \\
    \big( \lambda_1 t_0 + t_1\big)
    \Big( y_0^3+y_1^3+y_2^3 \Big) 
    +
    \big( \lambda_2 t_0 + \lambda_3 t_1\big)
    \Big( y_0 y_1 y_2 \Big)
    = 0
  \end{array}
  ,
\end{equation}
where we labelled three particular complex structure moduli $\lambda_1,
\lambda_2, \lambda_3\in \C$. The significance of these three moduli is
that they parametrize $\ZZZ$ symmetric complete
intersections. Explicitly, the 
group action is ($\zeta=e^{\frac{2\pi i}{3}}$)
\begin{equation}
  \begin{split}
    g_1:&\;
    \begin{cases}
      [x_0:x_1:x_2] \mapsto
      [x_0:\zeta x_1:\zeta^2 x_2]
      \\
      [t_0:t_1] \mapsto
      [t_0:t_1] 
      ~\text{(no action)}
      \\
      [y_0:y_1:y_2] \mapsto
      [y_0:\zeta y_1:\zeta^2 y_2]
    \end{cases}
    \\
    g_2:&\;
    \begin{cases}
      [x_0:x_1:x_2] \mapsto
      [x_1:x_2:x_0]
      \\
      [t_0:t_1] \mapsto
      [t_0:t_1] 
      ~\text{(no action)}
      \\
      [y_0:y_1:y_2] \mapsto
      [y_1:y_2:y_0]
      .
    \end{cases}
  \end{split}
\end{equation}
Since the two equations are each cubic in one $\CP^2$, one can
easily see that $\Xt$ is a double elliptic fibration. Hence its Euler
number must be $\chi(\Xt)=\chi(X)=0$. A straightforward geometric
calculation yields the Hodge diamond, which turns out to be
\begin{equation}
  \label{eq:HodgeDiamond}
  h^{p,q}\big(\Xt\big) = ~
  \vcenter{\xymatrix@!0@=7mm@ur{
      1 &  0  &  0  & 1 \\
      0 &  19 &  19 & 0 \\
      0 &  19 &  19 & 0 \\
      1 &  0  &  0  & 1
      ,
    }}
  \quad
  h^{p,q}\big(X\big) = ~
  \vcenter{\xymatrix@!0@=7mm@ur{
      1 &  0  &  0  & 1 \\
      0 &  3  &  3  & 0 \\
      0 &  3  &  3  & 0 \\
      1 &  0  &  0  & 1
    }}
\end{equation}
Note that, as mentioned above, $X$ has $h^{2,1}(X)=3$ complex
structure parameters.

\subsection{Coinvariant Homology}
\label{sec:coinv}

The Hurewicz isomorphism tells us that 
\begin{equation}
  H_1(X,\Z) = \pi_1(X)_\text{ab} = \Z_3 \oplus \Z_3
  .
\end{equation}
Together with the Hodge diamond eq.~\eqref{eq:HodgeDiamond}, the
Universal Coefficient Theorem and Poincar\'e duality, this determines
the integral homology up to a finite, Abelian group $T$
(``torsion''). The resulting integral homology groups are given in
Table~\ref{tab:HXZ}. 
\begin{table}[h]
\label{tab:HXZ}
\renewcommand\arraystretch{1.5}
\begin{displaymath}
\begin{array}{|c||c|c|c|c|c|c|c|}
\hline
i & 
0 & 1 & 2 & 3 & 4 & 5 & 6
\\ \hline
H_i(X,\Z) &
\Z & \Z_3 \oplus \Z_3  
& \Z^{3} \oplus T & \Z^{8} \oplus T & \Z^{3}\oplus \Z_3 \oplus \Z_3  & 0
& \Z
\\ \hline
H^i(X,\Z) &
\Z & 0
& \Z^{3} \oplus \Z_3 \oplus \Z_3  & \Z^{8} \oplus T & \Z^{3}\oplus T &
\Z_3 \oplus \Z_3  & \Z
\\ \hline
\end{array}
\end{displaymath}
\caption{Integral homology and cohomology of $X$.}\label{eqtable}
\end{table}
The hard part is to compute the torsion part
$T=H_2(X,\Z)_\text{tors}$, that is, the torsion homology classes of
curves in $X$. Clearly, such a torsion homology class cannot be
represented by a holomorphic curve. However, as we will see, torsion
homology classes can be represented by formal differences of
holomorphic curves. It turns out that 
\begin{equation}
  \label{eq:H2Xresult}
  H_2(X,\Z) = \Z^3 \oplus \underbrace{\Z_3 \oplus \Z_3}_{=T}
\end{equation}
How did we obtain this? First, note that by definition we have a
quotient map $q$ which induces the push-forward in homology,
\begin{equation}
  \vcenter{\xymatrix{
      \Xt \ar[d]_{q}
      & 
      H_2(\Xt,\Z)=\Z^{19}
      \ar[d]_{q_\ast}
      \\
      X
      & 
      H_2(X,\Z)=\Z^3\oplus \Z_3 \oplus \Z_3 
    }}
\end{equation}
The $\ZZZ$ group action on $\Xt$ identifies curves 
\begin{equation}
  \Ct = g \Ct 
  \qquad
  \forall g \in \ZZZ ,~ \Ct \in H_2(\Xt,\Z)
  .
\end{equation}
The quotient under this equivalence relation is called ``coinvariant
Homology''\footnote{Since this is the dual of invariant Cohomology.},
and denoted as $H_2\big(\Xt,\Z\big)_{\ZZZ}$. By explicitly
constructing a basis of curves on $X$ and identifying the group action
on them, one can explicitly determine the $\ZZZ$-group action on
$H_2\big(\Xt,\Z\big)\simeq \Z^{19}$ in terms of two commuting
$19\times 19$ matrices. We found that~\cite{Braun:2007xh}
\begin{equation}
  \begin{split}
    H_2\big(\Xt,\Z\big)_{\ZZZ} =&\; 
    H_2\big(\Xt,\Z\big) \Big/ 
    \mathop{\text{Span}}\big\{ \Ct - g \Ct \big\} 
    = 
    \\ 
    =&\;
    \Z^3 \oplus \Z_3 \oplus \Z_3
    .
  \end{split}
\end{equation}
In general, the relationship between coinvariant homology and the
homology of the quotient $X=\Xt/(\ZZZ)$ is determined by the
Cartan-Leray spectral sequence
\begin{equation}
  E^2_{p,q} =
  H_p\Big( \ZZZ, H_q\big(\Xt,\Z\big) \Big)
  \quad \Rightarrow \quad
  H_{p+q}\big(X,\Z\big)
  .
\end{equation}
Without going into details, we point out that the $p=0$ column of the
$E^2$ tableau is the coinvariant homology,
\begin{equation}
  E^2_{0,q} =  
  H_0\Big( \ZZZ, H_q\big(\Xt,\Z\big) \Big) =
  H_q\big(\Xt,\Z\big)_{\ZZZ}        
  .
\end{equation}
But, as usual, there are more entries in the spectral sequence, as
well as higher differentials and extension ambiguities. However, one
can show~\cite{Braun:2007xh} that these ultimately do not contribute
to $H_2(X,\Z)$.

To summarize, the homology classes of curves are given in
eq.~\eqref{eq:H2Xresult}. The torsion curve classes can be understood
as coinvariant curve classes on the covering space $\Xt$.

\section{Counting Curves}
\label{sec:Counting}

\subsection{Prepotential}
\label{sec:Prepotential}

Having constructed the Calabi-Yau threefold $X$, we would like to
identify the holomorphic curves on it. The well-known trick to count
the curves is via the instanton corrections to the Yukawa couplings in
the standard embedding,
\begin{equation}
  W_\text{Yuk}^\text{np}
  \big(t,\Rep{27}\big)
  =
  \sum_{\vec{n}\in H_2(X) }
  \Rep{27}_i
  \Rep{27}_j
  \Rep{27}_k
  \underbrace{
    C_{ijk}(\vec{n})
  }_{\in \Q}
  \prod q_a^{n_a}
\end{equation}
The instanton numbers are encoded in the power series 
$\sum_{\vec{n}} C_{ijk}(\vec{n}) \prod q_a^{n_a}$, multi-indexed by
$i,j,k=1,\dots,h^{1,1}$. Fortunately, the $(h^{1,1})^3$ power series
come from a single prepotential
\begin{equation}
  \label{eq:Prepotential}
  \FprepotNP{X} (t) = 
  \sum_{\vec{n}\in H_2(X) }
  \underbrace{
    n_{\vec{n}} 
  }_{\in \Z}
  \Li_3\left(
    \prod q_a^{n_a}
  \right)
  .
\end{equation}
The trilogarithm in the prepotential takes care of multi-covers of a
single instanton. Written as in eq.~\eqref{eq:Prepotential}, the
expansion coefficients $n_{\vec{n}}$ are the instanton numbers, and
they count\footnote{If all curves are rigid, then the instanton number
  is the naive number of curves in the given homology
  class. Otherwise, the expansion coefficients $n_{\vec{n}}$ are still
  integers, but may be negative.} the holomorphic curves in the
homology class $\vec{n}\in H_2(X,\Z)$.

\subsection{Torsion Curves}
\label{sec:prepot_intro}

In our case, the curve homology classes are
\begin{equation}
  H_2\big(X,\Z\big) = H_2\big(\Xt,\Z\big)_\ZZZ
  = \Z^3 \oplus \Z_3 \oplus \Z_3
  ,
\end{equation}
and we have to take into account the finite subgroup. Therefore, the
prepotential $\FprepotXNP$ must be
\begin{itemize}
\item a power series in the $h^{11}(X)$ variables $p,q,r$, and
\item a polynomial in $b_1,b_2$ with $b_i^3=1$.
\end{itemize}
After properly distinguishing the torsion curves, the prepotential
must be of the form
\begin{equation}
  \FprepotXNP
  = \sum_{\substack{\scriptstyle (n_1,n_2,n_3,m_1,m_2)\\ \in H_2(X,\Z)}}
  n_{(n_1,n_2,n_3,m_1,m_2)} 
  \Li_3 \big( p^{n_1} q^{n_2} r^{n_3} b_1^{m_1} b_2^{m_2} \big)
  .
\end{equation}
Knowing the general form of the series expansion, how do we go about
and compute it? We compared three different
approaches~\cite{Braun:2007xh, Braun:2007vy}
\begin{itemize}
\item Start on $\Xt$ and mod out $\ZZZ$.\\
  The prepotential on $\Xt$ is partially
  known~\cite{Hosono:1997hp}. The quotienting by the $\ZZZ$ action
  essentially amounts to a variable substitution~\cite{Braun:2007xh},
  and one obtains
  \begin{equation}
    \label{eq:F1}
    \begin{split}
      \FprepotXNP 
      & 
      (p,q,r, b_1,b_2) 
      = \frac{1}{9} p \,
      \times 
      \\ &\;
      \renewcommand{\arraystretch}{1.3}
      \begin{array}[t]{l}
      \ThetaEeight\big( 
      q^3;\, 
      q^2 b_1^2 b_2^2,\, q b_2,\, 1,\, b_1^2 b_2,\, 
      b_2^2,\, 1,\, b_1^2,\, q^{-1} \big) 
      \times
      \\ 
      \ThetaEeight\big( 
      r^3;\, 
      r^2 b_1 b_2,\, r b_2^2,\, 1,\, b_1 b_2^2,\, 
      b_2,\, 1,\, b_1,\, r^{-1} \big) 
      \times
      \\
      P\big( q^3 \big)^{12}    
      P\big( r^3 \big)^{12}            
      + O(p^2)
      .
      \end{array}
    \end{split}
  \end{equation}
\item Identify and count curves directly on $X$.\\
  By counting sections and singular Kodaira fibers in the double
  torus fibration $X$, one can directly count the curves up to linear
  order in $p$. The result is 
  \begin{equation}
    \label{eq:F2}
    \FprepotXNP 
    (p,q,r, b_1,b_2) 
    =
    p
    \left( \sum_{i,j=0}^2 b_1^i b_2^j \right)
    P(q)^4 P(r)^4 + O(p^2)
    .
  \end{equation}
  Setting $b_1=b_2=1$, the equality of eq.~\eqref{eq:F1} and
  eq.~\eqref{eq:F2} is a known identity of $E_8$ theta
  functions~\cite{MR1672085}, but we do not know a mathematical proof
  in general.
\item Toric mirror symmetry.\\
  This approach yields the full power series for the whole
  prepotential, but only numerical with degree bounded by computer
  resources. In practice, it is useful to look at (toric)
  $Z_3$-quotients of the toric complete intersection
  $\Xt$~\cite{Braun:2007vy}.
\end{itemize}

\subsection{Instanton Numbers}
\label{sec:numbers}

The instanton numbers $n_{(1,n_2,n_3,m_1,m_2)}$ turn out to be
independent of $m_1$ and $m_2$, that is, the same for all 
\begin{table}[htbp]
  \centering
  \renewcommand{\arraystretch}{1}
  \newcommand{\s}{\scriptstyle}
  \begin{tabular}{c|cccccccccc}
    \backslashbox{$\mathrlap{n_2}$}{$\mathclap{n_3~}$}
    &
    $0$ & $1$ & $2$ & $3$ & $4$ & $5$ & $6$ & $7$ 
    \\ \hline
    $0$ &
    $1$&$4$&$14$&$40$&$105$&$252$&$574$&$\s1240$
    \\
    $1$ &
    $4$&$16$&$56$&$160$&$420$&$\s1008$&$\s2296$&$\s4960$
    \\
    $2$ &
    $14$&$56$&$196$&$560$&$\s1470$&$\s3528$&$\s8036$&$\s17360$
    \\
    $3$ &
    $40$&$160$&$560$&$\s1600$&$\s4200$&$\s10080$&$\s22960$&$\s49600$
    \\
    $4$ &
    $105$&$420$&$\s1470$&$\s4200$&$\s11025$&$\s26460$&$\s60270$&$\s130200$
    \\
    $5$ &
    $252$&$\s1008$&$\s3528$&$\s10080$&$\s26460$&$\s63504$&$\s144648$&$\s312480$
    \\
    $6$ &
    $574$&$\s2296$&$\s8036$&$\s22960$&$\s60270$&$\s144648$&$\s329476$&$\s711760$
    \\
    $7$ &
    $\s1240$&$\s4960$&$\s17360$&$\s49600$&$\s130200$&$\s312480$&$\s711760$&$\s1537600$
  \end{tabular}
  \medskip 
  \caption{The instanton numbers $n_{(1,n_2,n_3,\ast,\ast)}$.}
  \label{tab:Numbers1}
\end{table}
$3\times 3=9$ choices of the torsion part. We list them in
Table~\ref{tab:Numbers1}. In particular, observe that
\begin{equation}
  n_{(1,0,0,\ast,\ast}) = 1
  ,
\end{equation}
that is, the $9$ smallest volume curves are alone in their homology
class. Hence, their contribution to the bundle moduli superpotential
cannot cancel within their curve class and Beasley-Witten's
cancellation cannot happen on the non-toric Calabi-Yau threefold $X$.

Although it is a priori surprising that the instanton numbers in
Table~\ref{tab:Numbers1} do not depend on the torsion part of the
homology class, this can be argued form a remaining discrete
symmetry. However, this phenomenon is special to $n_1=1$ and does not
hold in general. For example, 
\begin{table}[htbp]
  \centering
  \newcommand{\mathemph}[1]{\mbox{\boldmath $#1$}}
  \renewcommand{\arraystretch}{1.3}
  \newcommand{\s}{\scriptstyle}
  \newcommand{\sss}{\hspace{5mm}}
  \begin{tabular}{@{\sss}c@{\sss}@{\hspace{1mm}}@{\sss}c@{\sss}}
    $n_{(3,n_2,n_3,0,0)}$ &
    $n_{(3,n_2,n_3,m_1,m_2)},~(m_1,m_2)\not=(0,0)$
    \\
    \begin{tabular}{c|ccc}
      \backslashbox{$\mathrlap{n_2}$}{$\mathclap{n_3~}$}
      &
      $0$ & $1$ & $2$ 
      \\ \hline
      $0$ &
      $0$&$\mathemph{3}$&$\mathemph{36}$
      \\
      $1$ &
      $\mathemph{3}$&$\mathemph{108}$
      \\
      $2$ &
      $\mathemph{36}$
    \end{tabular}
    &
    \begin{tabular}{c|ccc}
      \backslashbox{$\mathrlap{n_2}$}{$\mathclap{n_3~}$}
      &
      $0$ & $1$ & $2$ 
      \\ \hline
      $0$ &
      $0$&$\mathemph{0}$&$\mathemph{27}$
      \\
      $1$ &
      $\mathemph{0}$&$\mathemph{81}$
      \\
      $2$ &
      $\mathemph{27}$
    \end{tabular}
  \end{tabular}
  \medskip 
  \caption{The instanton numbers $n_{(3,n_2,n_3,\ast,\ast)}$.}
  \label{tab:Numbers3}
\end{table}
the instanton numbers
$n_{(3,n_2,n_3,m_1,m_2)}$ listed in Table~\ref{tab:Numbers3} do
explicitly depend on $m_1$ and $m_2$.

\bibliographystyle{amsalpha}

\end{document}